\documentclass[runningheads]{llncs}

\usepackage{amsmath}
\usepackage{amssymb}
\usepackage{color}
\usepackage{graphicx}
\usepackage{hyperref}

\usepackage{xspace}

\newcommand{\desc}{\ensuremath{\mathtt{Desc}}\xspace}
\newcommand{\prop}{\ensuremath{\mathtt{prop}}\xspace}
\newcommand{\val}{\ensuremath{\mathtt{val}}\xspace}
\newcommand{\score}{\ensuremath{\mathtt{score}}\xspace}

\begin{document}

\title{DeepLENS: Deep Learning for Entity Summarization}

\author{Qingxia Liu \and
Gong Cheng \and
Yuzhong Qu}
\authorrunning{Q. Liu et al.}
\institute{
National Key Laboratory for Novel Software Technology, Nanjing University, China\\
\email{qxliu2013@smail.nju.edu.cn, \{gcheng,yzqu\}@nju.edu.cn}
}

\maketitle

\begin{abstract}
Entity summarization has been a prominent task over knowledge graphs. While existing methods are mainly unsupervised, we present DeepLENS, a simple yet effective deep learning model where we exploit textual semantics for encoding triples and we score each candidate triple based on its interdependence on other triples. DeepLENS significantly outperformed existing methods on a public benchmark.

\end{abstract}

\section{Introduction}

Entity summarization is the task of computing a compact summary for an entity by selecting an optimal size-constrained subset of entity-property-value triples from a knowledge graph such as an RDF graph~\cite{survey}. It has found a wide variety of applications, for example, to generate a compact entity card from Google's Knowledge Graph where an entity may be described in dozens or hundreds of triples. Generating entity summaries for general purposes has attracted much research attention, but existing methods are mainly unsupervised~\cite{relin,diversum,faces,facese,cd,linksum,bafrec,kafca,mpsum}. One research question that naturally arises is \emph{whether deep learning can much better solve this task}.

To the best of our knowledge, ESA~\cite{esa} is the only supervised method in the literature for this task. ESA encodes triples using graph embedding (TransE), and employs BiLSTM with supervised attention mechanism. Although it outperformed unsupervised methods, the improvement reported in~\cite{esa} was rather marginal, around~$+7\%$ compared with unsupervised FACES-E~\cite{facese} on the ESBM benchmark~\cite{esbm}. It inspired us to explore more effective deep learning models for the task of general-purpose entity summarization.

In this short paper, we present DeepLENS,\footnote{\url{https://github.com/nju-websoft/DeepLENS}} a novel \textbf{Deep} \textbf{L}earning based approach to \textbf{EN}tity \textbf{S}ummarization. DeepLENS uses a simple yet effective model which addresses the following two limitations of ESA, and thus achieved significantly better results in the experiments.
\begin{enumerate}
    \item Different from ESA which encodes a triple using graph embedding, we use word embedding because we consider textual semantics more useful than graph structure for the entity summarization task.
    \item Whereas ESA encodes a set of triples as a sequence and its performance is sensitive to the chosen order, our aggregation-based representation satisfies permutation invariance and hence more suitable for entity summarization.
\end{enumerate}

In the remainder of the paper, Section~\ref{sec:approach} details DeepLENS, Section~\ref{sec:experiments} presents experiment results, and Section~\ref{sec:conclusion} concludes the paper.
\section{Approach}\label{sec:approach}

\subsubsection{Problem Statement}

An RDF graph~$T$ is a set of triples. The \emph{description} of entity~$e$ in~$T$, denoted by~$\desc(e) \subseteq T$, comprises triples where $e$~is the subject or object. Each triple $t \in \desc(e)$ describes a property~$\prop(t)$ which is the predicate of~$t$, and gives a value~$\val(t)$ which is the object or subject of~$t$ other than~$e$. For a size constraint~$k$, a \emph{summary} of~$e$ is a subset of triples $S \subseteq \desc(e)$ with $|S| \leq k$. We aim to generate an optimal summary for general purposes.






\subsubsection{Overview of DeepLENS}

\begin{figure}[!t]
    \centering
    \includegraphics[width=0.75\linewidth]{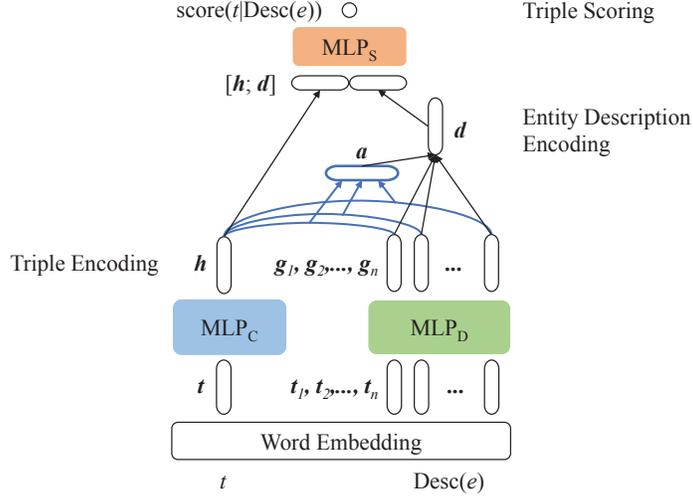}
    \caption{Model of DeepLENS.}
    \label{fig:model}
\end{figure}

Our approach DeepLENS generates an optimal summary by selecting $k$~most salient triples. As a supervised approach, it learns salience from labeled entity summaries. However, two issues remain unsolved. First, knowledge graph like RDF graph is a mixture of graph structure and textual content. The effectiveness of a learning-based approach to entity summarization relies on a \emph{proper representation of entity descriptions of such mixed nature}. Second, the salience of a triple is not absolute but dependent on the context, i.e.,~the set of other triples in the entity description. It is essential to \emph{represent their independence}. DeepLENS addresses these issues with the scoring model presented in Fig.~\ref{fig:model}. It has three modules which we will detail below: triple encoding, entity description encoding, and triple scoring. Finally, the model scores each candidate triple $t \in \desc(e)$ in the context of~$\desc(e)$.



\subsubsection{Triple Encoding}

For entity~$e$, a triple $t \in \desc(e)$ provides a property-value pair $\langle \prop(t),\val(t) \rangle$ of~$e$. Previous research~\cite{esa} leverages graph embedding to encode the structural features of~$\prop(t)$ and~$\val(t)$. By contrast, for the task of entity summarization we consider textual semantics more important than graph structure, and we \emph{solely exploit textual semantics} for encoding~$t$.


Specifically, for RDF resource~$r$, we obtain its \emph{textual form} as follows. For an IRI or a blank node, we retrieve its \texttt{rdfs:label} if it is available, otherwise we have to use its local name; for a literal, we take its lexical form. We represent each word in the textual form by a pre-trained word embedding vector, and we average these vectors over all the words to represent~$r$, denoted by $\text{Embedding}(r)$. For triple $t \in \desc(e)$, we generate and concatenate such vector representations for~$\prop(t)$ and~$\val(t)$ to form~$\boldsymbol{t}$, the \emph{initial representation} of~$t$. Then $\boldsymbol{t}$~is fed into a multi-layer perceptron (MLP) to generate~$\boldsymbol{h}$, the \emph{final representation} of~$t$:
\begin{equation}\label{eq:initial}
    \boldsymbol{t} = \left[\text{Embedding}(\prop(t)); ~\text{Embedding}(\val(t))\right] \,,\quad
    \boldsymbol{h} = \text{MLP}_\text{C}(\boldsymbol{t}) \,.\\
\end{equation}

\subsubsection{Entity Description Encoding}

To score a candidate triple in the context of other triples in the entity description, previous research~\cite{esa} captures the independence between triples in~$\desc(e)$ using BiLSTM to pass information. Triples are fed into BiLSTM as a sequence. However, $\desc(e)$~is a set and the triples lack a natural order. The performance of this model is unfavourably sensitive to the order of input triples. Indeed, as we will show in the experiments, different orders could lead to considerably different performance.


To generate a representation for~$\desc(e)$ that is \emph{permutation invariant}, we perform aggregation. Specifically, let $\boldsymbol{t_1}, \ldots, \boldsymbol{t_n}$ be the initial representations of triples in~$\desc(e)$ computed by Eq.~(\ref{eq:initial}). We feed a MLP with each~$\boldsymbol{t_i}$ for $1 \leq i \leq n$ and generate their final representations $\boldsymbol{g_1}, \ldots, \boldsymbol{g_n}$, which in turn are weighted using attention mechanism from~$\boldsymbol{h}$ computed by Eq.~(\ref{eq:initial}), the final representation of the candidate triple~$t$ to be scored. We calculate the sum of these weighted representations of triples to represent~$\desc(e)$, denoted by~$\boldsymbol{d}$:
\begin{equation}
    \boldsymbol{g_i} = \text{MLP}_\text{D}(\boldsymbol{t_i}) \,,\quad
    a_i = \frac{\exp(\cos(\boldsymbol{h}, \boldsymbol{g_i}))}{\sum_j \exp(\cos(\boldsymbol{h}, \boldsymbol{g_j}))} \,,\quad
    \boldsymbol{d} = \sum_{i=1}^{n}{a_i \boldsymbol{g_i}} \,.\\
\end{equation}
\noindent The result of summation is not sensitive to the order of triples in~$\desc(e)$.

\subsubsection{Triple Scoring}

For each candidate triple $t \in \desc(e)$ to be scored, we concatenate its final representation~$\boldsymbol{h}$ and the representation~$\boldsymbol{d}$ for~$\desc(e)$. We feed the result into a MLP to compute the context-based salience score of~$t$:
\begin{equation}
    \score(t|\desc(e)) = \text{MLP}_\text{S}(\left[\boldsymbol{h}; ~\boldsymbol{d}\right]) \,.
\end{equation}

Parameters of the entire model are jointly trained based on the mean squared error loss, supervised by labeled entity summaries.

\section{Experiments}\label{sec:experiments}

\subsection{Datasets} 

We used ESBM~v1.2, the largest available benchmark for evaluating general-purpose entity summarization.\footnote{\url{https://w3id.org/esbm}} For each of 125~entities in DBpedia and 50~entities in LinkedMDB, this benchmark provided 6~ground-truth summaries created by different human experts under $k=5$, and another 6~ground-truth summaries under $k=10$. We used the train-valid-test split specified in the benchmark to perform five-fold cross-validation.

\subsection{Participating Methods}

We compared DeepLENS with 10~baseline methods.

\textbf{Unsupervised Methods.}
We compared with 9~unsupervised methods that had been tested on ESBM: RELIN~\cite{relin}, DIVERSUM~\cite{diversum}, FACES~\cite{faces}, FACES-E~\cite{facese}, CD~\cite{cd}, LinkSUM~\cite{linksum}, BAFREC~\cite{bafrec}, KAFCA~\cite{kafca}, and MPSUM~\cite{mpsum}. We directly presented their results reported on the ESBM website.

\textbf{Supervised Methods.}
We compared with ESA~\cite{esa}, the only supervised method in the literature to our knowledge. We reused its open-source implementation and configuration.\footnote{\url{https://github.com/WeiDongjunGabriel/ESA}} We fed it with triples sorted in alphabetical order.

For our approach DeepLENS, we used 300-dimensional fastText~\cite{fasttext} word embedding vectors trained on Wikipedia to generate initial representations of triples. The numbers of hidden units in $\text{MLP}_\text{C}$, $\text{MLP}_\text{D}$, and $\text{MLP}_\text{S}$ were [64,~64], [64,~64], and [64,~64,~64], respectively. All hidden layers used ReLU as activation function. The final output layer of $\text{MLP}_\text{S}$ consisted of one linear unit. We trained the model using Adam optimizer with learning rate~0.01.

For both ESA and DeepLENS, we performed early stopping on the validation set to choose the number of training epochs from~1--50.

\textbf{Oracle Method.}
ORACLE approximated the best possible performance on ESBM and formed a reference point used for comparisons. It outputted $k$~triples that most frequently appeared in ground-truth summaries.

\subsection{Results}

Following ESBM, we compared machine-generated summaries with ground-truth summaries by calculating F1 score, and reported the mean F1 achieved by each method over all the test entities in a dataset.

\begin{table}[!t]
	\caption{Average F1 over all the test entities. Significant and insignificant differences ($p<0.01$) between DeepLENS and each baseline are indicated by~$\blacktriangle$ and~$\circ$, respectively.} 
	\label{tab:f1}
	\centering
	\resizebox{\textwidth}{!}{
	\begin{tabular}{|l|l|l|l|l|}
		\hline
		& \multicolumn{2}{|c|}{DBpedia} & \multicolumn{2}{|c|}{LinkedMDB} \\
		\cline{2-5}
		& $k=5$ & $k=10$ & $k=5$ & $k=10$ \\
		\hline

RELIN~\cite{relin}
 & 0.242
 & 0.455
 & 0.203
 & 0.258
 \\
DIVERSUM~\cite{diversum}
 & 0.249
 & 0.507
 & 0.207
 & 0.358
 \\
FACES~\cite{faces}
 & 0.270
 & 0.428
 & 0.169
 & 0.263
 \\
FACES-E~\cite{facese}
 & 0.280
 & 0.488
 & 0.313
 & 0.393
 \\
CD~\cite{cd}
 & 0.283
 & 0.513
 & 0.217
 & 0.331
 \\
LinkSUM~\cite{linksum}
 & 0.287
 & 0.486
 & 0.140
 & 0.279
 \\
BAFREC~\cite{bafrec}
 & 0.335
 & 0.503
 & 0.360		
 & 0.402
 \\
KAFCA~\cite{kafca}
 & 0.314		
 & 0.509
 & 0.244
 & 0.397		
 \\
MPSUM~\cite{mpsum}
 & 0.314		
 & 0.512
 & 0.272
 & 0.423
 \\
\hline
ESA~\cite{esa}
 & 0.331 
 & 0.532 
 & 0.350 
 & 0.416 
 \\
DeepLENS
 & 0.402 \tiny{$^\blacktriangle$}\tiny{$^\blacktriangle$}\tiny{$^\blacktriangle$}\tiny{$^\blacktriangle$}\tiny{$^\blacktriangle$}\tiny{$^\blacktriangle$}\tiny{$^\blacktriangle$}\tiny{$^\blacktriangle$}\tiny{$^\blacktriangle$}\tiny{$^\blacktriangle$}
 & 0.574 \tiny{$^\blacktriangle$}\tiny{$^\blacktriangle$}\tiny{$^\blacktriangle$}\tiny{$^\blacktriangle$}\tiny{$^\blacktriangle$}\tiny{$^\blacktriangle$}\tiny{$^\blacktriangle$}\tiny{$^\blacktriangle$}\tiny{$^\blacktriangle$}\tiny{$^\blacktriangle$}
 & 0.474 \tiny{$^\blacktriangle$}\tiny{$^\blacktriangle$}\tiny{$^\blacktriangle$}\tiny{$^\blacktriangle$}\tiny{$^\blacktriangle$}\tiny{$^\blacktriangle$}\tiny{$^\blacktriangle$}\tiny{$^\blacktriangle$}\tiny{$^\blacktriangle$}\tiny{$^\blacktriangle$}
 & 0.493 \tiny{$^\blacktriangle$}\tiny{$^\blacktriangle$}\tiny{$^\blacktriangle$}\tiny{$^\blacktriangle$}\tiny{$^\blacktriangle$}\tiny{$^\blacktriangle$}\tiny{$^\blacktriangle$}\tiny{$^\blacktriangle$}\tiny{$^\blacktriangle$}\tiny{$^\blacktriangle$}
 \\
\hline
ORACLE
 & 0.595 & 0.713 & 0.619 & 0.678 \\
    \hline
	\end{tabular}
	}
\end{table}

\textbf{Comparison with Baselines.}
As shown in Table~\ref{tab:f1}, supervised methods were generally better than unsupervised methods. Our DeepLENS outperformed all the baselines including ESA. Moreover, two-tailed t-test showed that all the differences were statistically significant ($p<0.01$) in all the settings. DeepLENS achieved new state-of-the-art results on the ESBM benchmark. However, the notable gaps between DeepLENS and ORACLE suggested room for improvement and were to be closed by future research.

\begin{table}[!t]
	\caption{Average F1 over all the test entities achieved by different variants of ESA.} 
	\label{tab:esa}
	\centering
	\begin{tabular}{|l|l|l|l|l|}
		\hline
		& \multicolumn{2}{|c|}{DBpedia} & \multicolumn{2}{|c|}{LinkedMDB} \\
		\cline{2-5}
		& $k=5$ & $k=10$ & $k=5$ & $k=10$ \\
		\hline
ESA
 & 0.331 
 & 0.532
 & 0.350
 & 0.416 \\
ESA-text
 & 0.379  
 & 0.558 
 & 0.390
 & 0.418 \\
ESA-rnd
 & 0.116\tiny{$\pm$0.008}
 & 0.222\tiny{$\pm$0.007}
 & 0.113\tiny{$\pm$0.015}
 & 0.219\tiny{$\pm$0.011} \\ 
 \hline
	\end{tabular}
\end{table}

\textbf{Ablation Study.}
Compared with ESA, we attributed the better performance of DeepLENS to two improvements in our implementation: the exploitation of textual semantics, and the permutation invariant representation of triple set. They were demonstrated by the following ablation study of ESA.

First, we compared two variants of ESA by encoding triples in different ways. For triple~$t$, the original version of ESA encoded the structural features of~$\prop(t)$ and~$\val(t)$ using TransE. We implemented ESA-text, a variant that encoded both~$\prop(t)$ and~$\val(t)$ using fastText as in our approach. As shown in Table~\ref{tab:esa}, ESA-text slightly outperformed ESA, showing the usefulness of textual semantics compared with graph structure used by ESA.

Second, we compared two variants of ESA by feeding with triples in different orders. The default version of ESA was fed with triples sorted in alphabetical order for both training and testing. We implemented ESA-rnd, a variant that was fed with triples in alphabetical order for training but in random order for testing.
We tested ESA-rnd 20~times and reported its mean F1 with standard deviation. In Table~\ref{tab:esa}, the notable falls from ESA to ESA-rnd showed the unfavourable sensitivity of BiLSTM used by ESA to the order of input triples.
\section{Conclusion}\label{sec:conclusion}

We presented DeepLENS, a simple yet effective deep learning model for general-purpose entity summarization. It has achieved new state-of-the-art results on the ESBM benchmark, significantly outperforming existing methods. Thus, entity summarization becomes another research field where a combination of deep learning and knowledge graph is likely to shine. However, in DeepLENS we only exploit textual semantics. In future work, we will incorporate ontological semantics into our model. We will also revisit the usefulness of structural semantics.

\section*{Acknowledgments}
This work was supported by the National Key R\&D Program of China under Grant 2018YFB1005100 and by the Qing Lan Program of Jiangsu Province.

\bibliographystyle{splncs04}
\bibliography{main}

\begin{thebibliography}{10}
\providecommand{\url}[1]{\texttt{#1}}
\providecommand{\urlprefix}{URL }
\providecommand{\doi}[1]{https://doi.org/#1}

\bibitem{fasttext}
Bojanowski, P., Grave, E., Joulin, A., Mikolov, T.: Enriching word vectors with
  subword information. {TACL}  \textbf{5},  135--146 (2017)

\bibitem{relin}
Cheng, G., Tran, T., Qu, Y.: {RELIN:} relatedness and informativeness-based
  centrality for entity summarization. In: {ISWC} 2011, Part {I}. pp. 114--129
  (2011)

\bibitem{faces}
Gunaratna, K., Thirunarayan, K., Sheth, A.P.: {FACES:} diversity-aware entity
  summarization using incremental hierarchical conceptual clustering. In:
  {AAAI} 2015. pp. 116--122 (2015)

\bibitem{facese}
Gunaratna, K., Thirunarayan, K., Sheth, A.P., Cheng, G.: Gleaning types for
  literals in {RDF} triples with application to entity summarization. In:
  {ESWC} 2016. pp. 85--100 (2016)

\bibitem{kafca}
Kim, E.K., Choi, K.S.: Entity summarization based on formal concept analysis.
  In: {EYRE} 2018 (2018)

\bibitem{bafrec}
Kroll, H., Nagel, D., Balke, W.T.: {BAFREC}: Balancing frequency and rarity for
  entity characterization in linked open data. In: {EYRE} 2018 (2018)

\bibitem{survey}
Liu, Q., Cheng, G., Gunaratna, K., Qu, Y.: Entity summarization: State of the
  art and future challenges. CoRR  \textbf{abs/1910.08252} (2019)

\bibitem{esbm}
Liu, Q., Cheng, G., Gunaratna, K., Qu, Y.: {ESBM}: An entity summarization
  benchmark. In: {ESWC} 2020 (2020)

\bibitem{diversum}
Sydow, M., Pikula, M., Schenkel, R.: The notion of diversity in graphical
  entity summarisation on semantic knowledge graphs. J. Intell. Inf. Syst.
  \textbf{41}(2),  109--149 (2013)

\bibitem{linksum}
Thalhammer, A., Lasierra, N., Rettinger, A.: {LinkSUM}: Using link analysis to
  summarize entity data. In: {ICWE} 2016. pp. 244--261 (2016)

\bibitem{mpsum}
Wei, D., Gao, S., Liu, Y., Liu, Z., Huang, L.: {MPSUM}: Entity summarization
  with predicate-based matching. In: {EYRE} 2018 (2018)

\bibitem{esa}
Wei, D., Liu, Y., Zhu, F., Zang, L., Zhou, W., Han, J., Hu, S.: {ESA}: Entity
  summarization with attention. In: {EYRE} 2019. pp. 40--44 (2019)

\bibitem{cd}
Xu, D., Zheng, L., Qu, Y.: {CD} at {ENSEC} 2016: Generating characteristic and
  diverse entity summaries. In: {SumPre} 2016 (2016)

\end{thebibliography}

\end{document}